\documentclass[superscriptaddress,twocolumn]{revtex4-1}

\usepackage[T1]{fontenc}
\usepackage{amsmath}
\usepackage{amssymb}
\usepackage{amsbsy}
\usepackage{minibox}
\usepackage{microtype}
\usepackage{etoolbox}
\usepackage{bm}
\usepackage[usenames,dvipsnames,svgnames,table]{xcolor}
\usepackage{graphicx}
\usepackage{textcomp}
\usepackage[english]{babel}
\usepackage{dcolumn} 
\usepackage{xspace}
\usepackage{braket}

\begin{document}

\title{A compact ultranarrow high-power laser system for experiments with $578$~nm Ytterbium clock transition}

\author{G. Cappellini}
\email{cappellini@lens.unifi.it}
\affiliation{LENS European Laboratory for Nonlinear Spectroscopy, 50019 Sesto Fiorentino, Italy}
\author{\textnormal{\textsuperscript{,b)}} P. Lombardi}
\thanks{These two authors contributed equally to this work.}
\affiliation{Department of Physics and Astronomy, University of Florence, 50019 Sesto Fiorentino, Italy}
\affiliation{INO-CNR Istituto Nazionale di Ottica del CNR, Sezione di Sesto Fiorentino, 50019 Sesto Fiorentino, Italy}
\author{M. Mancini}
\affiliation{Department of Physics and Astronomy, University of Florence, 50019 Sesto Fiorentino, Italy}
\author{G. Pagano}
\affiliation{Department of Physics and Astronomy, University of Florence, 50019 Sesto Fiorentino, Italy}
\author{M. Pizzocaro}
\affiliation{INRIM Istituto Nazionale di Ricerca Metrologica, Torino 10135, Italy}
\author{L. Fallani}
\affiliation{LENS European Laboratory for Nonlinear Spectroscopy, 50019 Sesto Fiorentino, Italy}
\affiliation{Department of Physics and Astronomy, University of Florence, 50019 Sesto Fiorentino, Italy}
\author{J. Catani}
\affiliation{LENS European Laboratory for Nonlinear Spectroscopy, 50019 Sesto Fiorentino, Italy}
\affiliation{INO-CNR Istituto Nazionale di Ottica del CNR, Sezione di Sesto Fiorentino, 50019 Sesto Fiorentino, Italy}

\begin{abstract}
In this paper we present the realization of a compact, high-power laser system able to excite the Ytterbium clock transition at $578$~nm. Starting from an external-cavity laser based on a quantum dot chip at $1156$~nm with an intra-cavity electro-optic modulator, we were able to obtain up to $60$~mW of visible light at $578$~nm  via frequency doubling. The laser is locked with a $500$~kHz bandwidth to a ultra-low-expansion glass cavity stabilized at its zero coefficient of thermal expansion temperature through an original thermal insulation and correction system. This laser allowed the observation of the clock transition in fermionic $^{173}$Yb with a $<50$~Hz linewidth over $5$ minutes, limited only by a residual frequency drift of some $0.1$~Hz/s. 
\end{abstract}

\maketitle

\section{INTRODUCTION}
The interest in ultranarrow laser sources has grown larger and larger in the recent years. The possibility to employ ultranarrow lasers to excite the so-called ``clock transitions'' of alkaline-earth and alkaline-earth-like atoms has proven to be a solid choice for the realization of new frequency standards \cite{clockoates,clockye,clockkatori} and also represents a fundamental resource for the implementation of new quantum simulators of fundamental physical models \cite{gorshkov2010} and for the realization of reliable quantum information schemes \cite{daley1,daley2}.\par
In this paper we describe the ultranarrow laser system at $578$~nm that we realized in order to excite the Ytterbium $^{1}S_0$ $\rightarrow$ $^{3}P_0$ clock transition \cite{fortson2005}, which recently allowed the investigation of a coherent spin-exchange dynamics between Ytterbium atoms in different electronic states \cite{nostroprl}. Several details regarding the design of the laser, the electronics and the insulation of the reference cavity will be given in this work.\par
Differently from other experiments, in which the light at $578$~nm is generated via Sum Frequency Generation \cite{uleoates,uleturin}, we employ a cavity-enhanced Second Harmonic Generation (SHG) scheme in order to obtain visible light at $578$~nm from the infrared (IR) radiation emitted by a solid state gain chip. While SHG-based schemes have already been adopted \cite{nevsky2008,lee2011}, in our system we were able to obtain a large amount of power ($60$~mW) of visible light. This represents a very important feature in order to observe and exploit the clock transition in Ytterbium bosonic isotopes where the $^{1}S_0$ $\rightarrow$ $^{3}P_0$ transition is strictly forbidden because of the absence of hyperfine structure (due to the zero nuclear spin), but can be induced via magnetic quenching with the $^{3}P_1$ state \cite{quenching}. The strength of this induced transition is weaker than in fermionic isotopes, so a large optical power is desirable. Moreover, such high power could allow for off-resonant manipulation of ultracold fermionic ytterbium, e.g. for the realization of spin selective optical dressing or for optical control of atom-atom scattering \cite{jones2006}.\par
The laser radiation frequency is stabilized to a thermally and acoustically insulated Ultra Low Expansion (ULE) glass cavity and the whole optical system and its electronics lay on a dedicated optical table that can be easily moved and transported.

\section{INFRARED LASER SOURCE} \label{sec:source}
\begin{figure}[t]
\centering
    \includegraphics[width=0.9\columnwidth]{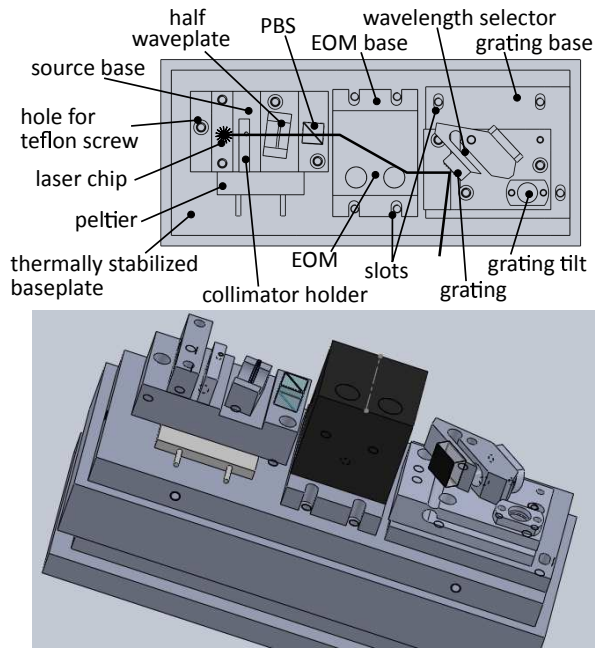}
\caption{Top and 3D view of the long cavity ECDL laser source at 1156 nm. Intracavity PBS and waveplate are tilted in order to avoid backreflections.}\label{fig:ECDL}
\end{figure}

The source of our laser system is a quantum dot gain chip (Innolume GC-1156-TO-200) powered by a ILX Lightwave LDX-3620 Ultra Low Noise current driver. The frequency-to-current response of this chip shows a continuous phase slip from $0$ to $180\text{\textdegree}$ between $10$ and $100$~kHz of current modulation frequency, making current a poor actuator for laser frequency locking to a stable reference with a $>100$~kHz bandwidth. In order to obtain a robust and high-bandwidth frequency stabilization, we chose a long external-cavity laser (ECDL) configuration with a broadband electro-optic modulator (EOM) as fast cavity length actuator \cite{longECDL}. An increased cavity length of $12$~cm brings the advantage of an intrinsic smaller linewidth (approximately $20$~kHz) compared to standard ECDLs (typically of the order of hundreds of kHz \cite{ECDL}), at the expense of a reduced mode-hopping-free interval. In our case this is not a drawback, as we are able to cover more than one Free Spectral Range (FSR) of the reference ULE cavity without mode-hops (see section~\ref{sec:ULE}). A schematic drawing of the ECDL is reported in Fig.~\ref{fig:ECDL}. The EOM, manifactured by Qubig GmbH (model EO-DC5M-BREWSTER), relies on a Brewster-cut (in order to avoid reflections and Etalon effect) Magnesium doped LiNbO$_3$ crystal, and shows a nearly flat modulation depth and phase response up to the MHz scale. The resulting laser frequency-to-voltage transduction is of the order of 1~MHz/V, hence no high voltage amplifier is required to drive the EOM and a standard servo amplifier output can be used with benefits to the total bandwidth of the system. The two floating electrodes of the EOM are connected in differential configuration to SMA-type connectors so that two independent modulation signals can be used to drive the EOM.\par
A $\lambda/2$ waveplate and a polarizing beamsplitter (PBS) are placed inside the external cavity in order to precisely tune and clean the polarization of the radiation. We have indeed observed a marked tendency of the ECDL to lase on different linear polarization modes with an angle of $\sim15\text{\textdegree}$ between the two. The origin of this behavior is still unknown, but we exclude temperature fluctuations and current surges as possible causes since the laser temperature is stabilized and the supply current is accurately filtered. These intra-cavity polarizing elements also minimize the Residual Amplitude Modulation (RAM) due to the mismatch between the polarization of the light and the axes of the EOM crystal.\par
The optical feedback is provided by a diffraction grating optimized for UV (Thorlabs GH13-12U) that reinjects $\sim$10\% of the light into the laser chip and forms a Littrow external cavity. A piezoelectric stack (Piezomechanik PSt $150/4/7$) enables the tuning of the laser frequency over a span of about $1$~GHz. In this configuration, the laser is able to deliver up to $200$~mW of radiation at $1156$~nm with a $500$~mA supply current, which is the maximum output of our current driver. The chip is actually capable to sustain supply currents up to $700$~mA in external cavity, further increasing the ECDL output power up to $250$~mW.\par
The quantum dot chip and the grating are mounted on holders machined from Ergal aluminium alloy screwed to a monolithic massive Anticorodal baseplate, allowing for precise collimation of the laser beam and reinjection of the first diffraction order. A home-made Arduino-based digital temperature controller stabilizes at $26.80\pm0.01$~\textdegree C the temperature of the laser baseplate (and of the ULE reference cavity, further details will be given in section \ref{sec:ULE}). Also the EOM is in direct thermal contact with the stabilized baseplate through an improved ceramic thermalization mount. A Peltier stage between the baseplate and the laser diode holder keeps the gain chip temperature fixed at $27$~\textdegree C with $1$~mK precision. Moreover, the ECDL is placed in a box providing thermal and mechanical insulation. The full temperature stabilization system limits the ECDL frequency drift within a few tens of MHz during one day of operation when not locked to the ULE cavity frequency reference.

\section{SECOND HARMONIC GENERATION}
In our setup, the visible light at $578$~nm is generated via SHG realized through a periodically poled LiNbO$_3$ nonlinear crystal placed into a symmetrical bow-tie optical cavity. The crystal, manufactured by HC Photonics, is doped with 5\% of MgO to raise its damage threshold and has a poling period of $8.9$~$\mu$m. A copper oven thermally stabilizes the crystal at the temperature of maximum conversion efficiency of $65$~\textdegree C and is mounted on a four-axis translation stage (New Focus Model 9071) to adjust the crystal position with respect to the cavity optical axis. \par
The bow-tie optical cavity consists of two concave mirror with $100$~mm radius of curvature and two plane mirrors, one of which mounted on a low capacity piezoelectric stack (Piezomechanik PSt $150/2$x$3/7$, $170$~nF) allowing the lock of the cavity length to the infrared laser source. The cavity has a $545$~MHz FSR and a finesse $\sim100$. The mirror holders are fixed to a monolithic Anticorodal base that rests upon four small $1/4$ inch-thick Sorbothane supports to decouple the whole cavity from acoustic noise delivered through the table top. The cavity is placed inside an aluminum enclosure that is evacuated to $10^{-2}$~bar to further reduce its sensitivity to air-delivered noise. The vacuum environment also suppresses bistability operation caused by a non-linear dispersive effect induced by the presence of water vapor resonances around 1156 nm.\par
\begin{figure*}[t]
\begin{center}
\includegraphics[width=\textwidth]{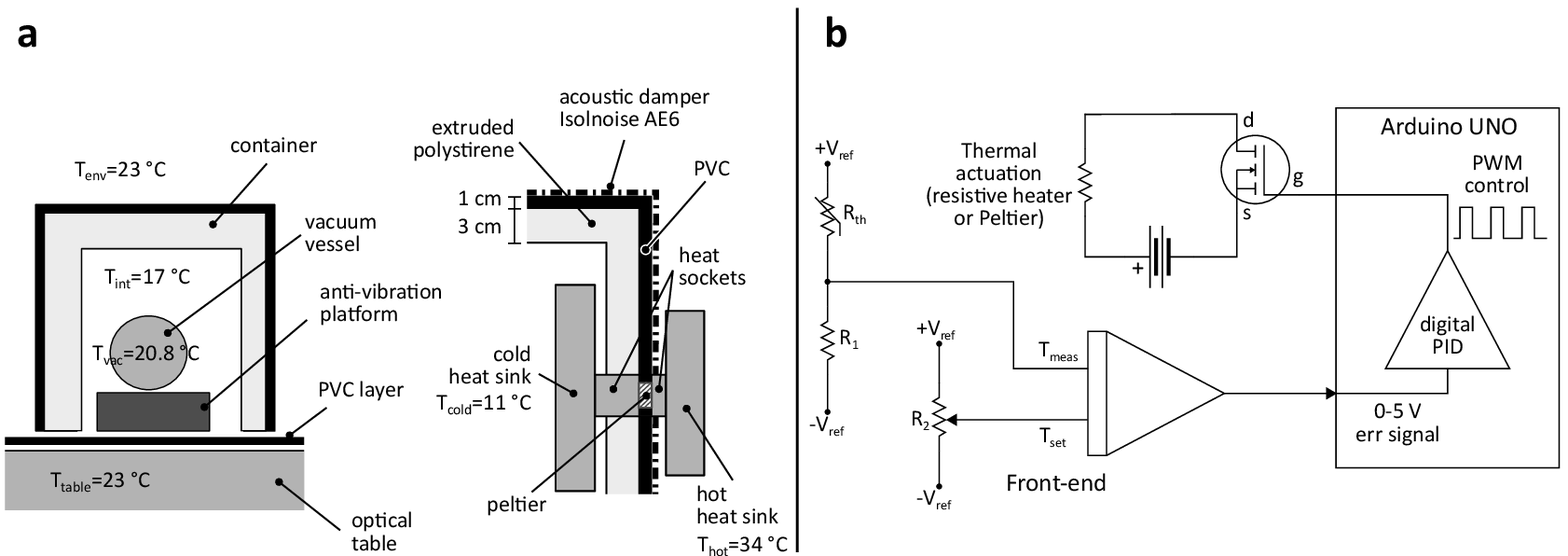}
\end{center}
\caption{{\bf a.} Sketch of the insulation system of the ULE cavity and detail of one of the four elements of the enclosure Peltier-based cooling stage. {\bf b.} Essential schematics of the home made digital temperature controller. The analog front-end measures the temperature reading the voltage of a divider between a thermistor ($R_{th}$) and a precision resistor ($R_1$) and compares it to the setpoint temperature (controlled by $R_2$). The error signal is translated into the $0-5$ V range and amplified so that it can be fed to the Arduino 10-bit ADC without loss of precision. Arduino computes and generates the correction at the PWM outputs and drives a MOSFET gate to regulate the mean current flow in a thermal actuation element (a set of resistors for the ULE cavity and laser baseplate or a Peltier for the enclosure, see text).}
\label{fig:fridge}
\end{figure*} 
The error signal used to keep the SHG cavity resonance locked to the IR laser frequency is obtained employing the H\"ansch-Couillaud technique \cite{HClock}. A single proportional-integral (PI) servo amplifier stage provides the feedback on the cavity piezo with a $-3$~dB bandwidth of $300$~Hz. With this system, we are able to obtain up to $60$~mW of $578$~nm light from $150$~mW of 1156~nm radiation ($40\%$ conversion efficiency), which to our knowledge is the highest power at this wavelength reported in literature. The majority of the produced visible light is delivered to the experiment table through a polarization maintaining (PM) optical fiber, while a small fraction of it ($\sim1$~mW) is sent to the reference ULE cavity for frequency stabilization of the laser source.

\section{ULE REFERENCE CAVITY} \label{sec:ULE} 
In order to reduce the laser linewidth to the Hz level, it is required to lock the source to a very stable reference. Following a well established procedure \cite{oatesMethods}, we lock our source to a high-finesse ULE glass Fabry-Perot resonator via the Pound-Drever-Hall (PDH) technique \cite{PDHlock}. With a proper PDH lock loop, the laser frequency noise spectrum substantially reproduces that of the reference cavity. For this reason, it is fundamental to minimize the cavity sensitivity to mechanical vibrations that would result in laser frequency noise at acoustic frequencies thus increasing the laser fast linewidth. In order to perform experiments on atomic clock transitions it is also crucial to greatly reduce any laser frequency drift. ULE glass is chosen because its coefficient of thermal expansion (CTE) has a zero around room temperature. It is hence important to thermally insulate the cavity and carefully determine its zero CTE temperature in order to minimize residual length fluctuations.\par 
The ULE system employed in our setup, manufactured by Advanced Thin Films, was fully characterized in a previous experiment \cite{uleturin}. Both the mirrors substrates and the spacer composing the reference cavity are fabricated in ULE glass. The cavity FSR is $1.5$~GHz and its finesse, measured by means of ring-down technique, results $\sim150000$. In order to minimize sensitivity to external vibrations, the cavity, shaped as a notched cylinder, is suspended horizontally on viton pads at its Airy points, over an aluminum holder. The position of these points has been previously determined with a finite element analysis \cite{uleturin}. For a better acoustic and thermal insulation, the cavity, surrounded by a cylindrical thermal copper shield, is positioned in a horizontal stainless steel $25$~cm long CF200 vacuum vessel and kept at $10^{-7}$~mbar by a $2$~L/s ion pump. The vacuum chamber is placed on a passive vibration-insulation platform (MinusK BM-8) which grants a reduction between $10$ and $20$~dB of the acoustic noise in the Fourier frequency range from $1$ to $100$~Hz \cite{uleturin}. As a final thermal and acoustic insulation stage from the lab environment, the whole system is located inside an $80$~cm~$\times$~$80$~cm~$\times$~$80$~cm home-made enclosure made of joint layers of PVC and extruded polystyrene. A sketch of the structure is reported in Fig.~\ref{fig:fridge}a. The whole enclosure has a relatively low weight ($\sim30$~kg). Extruded polystyrene is one of the best commercially available thermal insulator (specific thermal resistivity $\lambda_{EP}^{-1}\sim 30$ m$\cdot$K/W), while PVC is able to provide the mechanical support for the structure, and is a fair thermal insulator ($\lambda_{PVC}^{-1}\sim 5$ m$\cdot$K/W). We also applied on the external surface of the enclosure a resilient mat (Aetolia Isolnoise AE6) as an additional low-frequency acoustic damper. We used commercial glue for linoleum floors to stick together all the plastic layers. The box can be opened to grant access to the optics needed to couple the light to the ULE cavity, including the PDH EOM and other polarization optics that benefit from a thermally stable environment in terms of lock performances (see section \ref{sec:lock}).\par
In order to stabilize the temperature of the ULE cavity (and of the ECDL baseplate, see section \ref{sec:source}) we developed a home-made multi-channel low-cost digital temperature controller based on Arduino UNO architecture. A scheme of the controller is reported in figure Fig.~\ref{fig:fridge}b. The front-end of every controller channel measures the temperature with a thermistor and compares it to the setpoint, generating an error signal centered around $0$~V. This error signal is amplified and offset by $2.5$~V in order to obtain a $0-5$~V signal that is fed to one of the Arduino 10 bit analog inputs. The amplification stage allows to ``zoom-in'' around the setpoint, overcoming the limitation represented by the Arduino ADC resolution so that the desired controller precision can be achieved simply tuning its gain. The algorithm running on Arduino computes a PI correction for every error signal and the outputs are generated at the pulse-width-modulation (PWM) pins. Every PWM channel controls the gate of a MOSFET transistor acting as a switch to regulate the current flow through a set of resistive heaters, allowing temperature stabilization with minimal power dissipation on the MOSFET transistor. At the moment, only two front-ends are used, but their number can be increased up to six. In order to stabilize the cavity temperature, the controller acts on two flat ribbon resistive heaters placed on the external surface of the vacuum chamber with an heating power of around $1.2$~W at regime. The system handles up to $2.5$~W during transients. The thermistor is placed on the copper shield and not directly on the cavity in order not to break its symmetry thus increasing vibration sensitivity. Current wires are tightly twisted in order to avoid irradiated noise at the PWM frequency (480 Hz). Due to the high vacuum environment, the ULE cavity thermalizes at the temperature of the copper shield via irradiation since all of the other heat exchange mechanisms are negligible.\par
The measured zero CTE temperature of our ULE cavity is of $20.80\pm0.03$~\textdegree C. This temperature is lower than our working environment ($23\pm1$~\textdegree C), so cooling is required in order to stabilize the cavity at its zero CTE temperature. An optimal Peltier cooling system would require two vacuum thermal shields for an ideal heat exchange \cite{ulehansch}, while in our system a single thermal shield is present. On the other hand, cooling the vacuum chamber with Peltier cells coupled to standard heat sinks would create thermal gradients due to the bad thermal conductivity of stainless steel. We also excluded the possibility to cool the vacuum vessel with liquid and air flows as this would induce mechanical vibration on the structure. The required cooling is provided by a low-cost Peltier-based heat transfer stage, consisting in a set of four Peltier elements (Global Component Sourcing ET-127-14-11) in thermal contact with internal and external heat sinks through aluminum sockets (Fig.~\ref{fig:fridge}a). The Peltier cells maintain the internal temperature around $17.5$~\textdegree C ($\sim5$~\textdegree C lower than lab temperature) stabilizing the internal heatsinks temperature at $12$~\textdegree C with a total power consumption of $20$~W. The effectiveness of this cooling stage is further increased with the addition of two slow speed fans ($\sim100$ rpm) on the external heat sinks with no visible effect on the laser system noise. Thermal stabilization of the internal heat sinks is implemented with a stand-alone digital temperature controller based on the scheme of Fig.~\ref{fig:fridge}b, which also adjusts its setpoint depending on the external temperature in order to compensate for different heat flows through the enclosure at different environmental temperatures.
This allows us to stabilize the inner air temperature below the $0.1$~\textdegree C level and keep it lower than the estimated zero CTE (see Fig.~\ref{fig:fridge}a). In this way the vacuum system can be stabilized around $21$~\textdegree C without the need of an in-vacuum Peltier. The thermalization time of the entire system (ULE cavity and external enclosure) is of the order of 2-3 days, still adequately manageable through the digital thermal stabilization system. \par
This scheme, in which both the reference cavity and its insulation enclosure are temperature controlled, could also allow for an easier stabilization of the cavity at high temperatures, heating the enclosure to an intermediate value between room temperature and the cavity target temperature.

\section{LASER FREQUENCY STABILIZATION} \label{sec:lock}
\begin{figure}[t]
\centering
	\includegraphics[width=0.9\columnwidth]{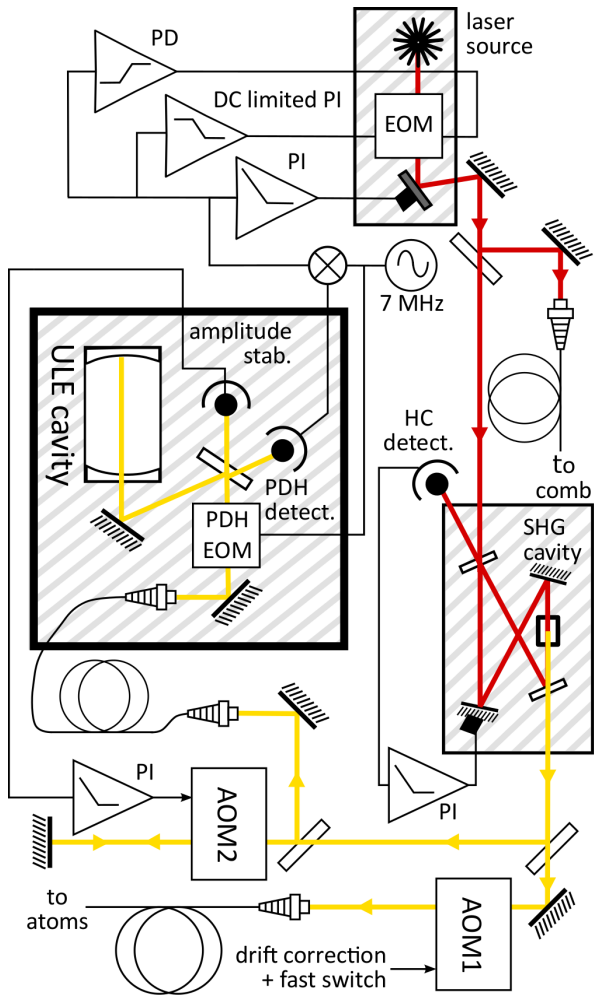}
\caption{Scheme of the optical setup and the locking loop electronics.}\label{fig:setup}
\end{figure}

The schematic of the laser system lock is reported in Fig.~\ref{fig:setup}. 
The fraction of visible light that is used for the frequency stabilization enters a double passage cat's eye path with an acousto-optic modulator (AOM1) operating at $105$~MHz and is then delivered inside the ULE insulation enclosure via a $2$~m long PM optical fiber. In order to minimize the RAM, the $7$~MHz EOM (Thorlabs EO-PM-NR-C4) used for the generation of the PDH signal is placed inside the thermal box together with a Glan-Thompson polarizer on a tilt mount used to carefully adjust the polarization of the incoming light. Moreover, in order to minimize multiple internal reflections that would result in a low frequency residual modulation of the PDH signal, the EOM is tilted with respect to the light propagation axis. A photodiode collects the incident light and drives a PI servo amplifier acting on AOM1 to stabilize the power at 60~$\mu$W in order to minimize power induced cavity resonance fluctuations (both photo-thermal and radiation pressure ones\cite{bergquist92}).\par
The PDH error signal is obtained by demodulating and low-pass filtering the output of a fast photodiode which collects the back-reflected light of the ULE glass cavity. The signal is fed in parallel to three servo amplifier stages which act on the ECDL piezo transducer and on the two intra-cavity EOM electrodes. The piezo servo works only in the very low frequency range (DC-$200$~Hz), and its transfer function is enhanced in the very low frequency range with an additional integrator stage. The intra-cavity EOM is driven by two independent servos, acting each on a single electrode with the ground as common reference. The first is a PI, with limited DC gain in order to avoid conflicts with the piezo loop. It provides corrections up to $200$~kHz. The second servo consists of a minimal circuit, AC coupled above $16$ kHz, with a single fast operational amplifier (AD818AN, $100$ MHz bandwidth at unity gain). It provides a proportional correction up to the MHz, with a derivative stage which helps in lowering the PI servo bump and shifts its frequency up to $\simeq0.5$~MHz. All the elements of the feedback loops are battery powered in order to reduce the electronic noise. We also carefully arranged a ``star-like'' topology for the ground circuit to minimize the noise induced by ground loops. \par
\begin{figure}[t]
\centering
    \includegraphics[width=1\columnwidth]{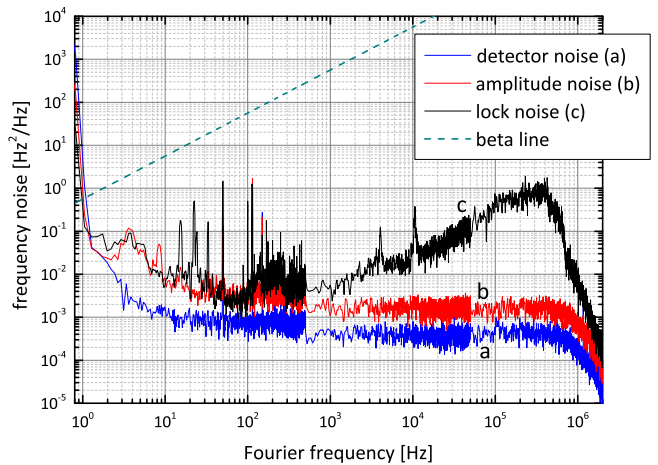}
\caption{Frequency noise spectrum of the locked signal compared to electronic noise and amplitude noise of the out of resonance error signal. The black trace (c) is the frequency noise of the laser locked to the ULE cavity. The red trace (b) is the amplitude noise of the completely out of resonance backreflected light. The blue trace (a) is the electronic noise of the photodiode and locking electronics. The Beta Line \cite{betaline} is a limit above which the frequency noise of the lock contributes to the laser linediwth.}\label{fig:spectrum}
\end{figure}  
As we do not have a second system to check the laser performances, we optimized the lock using for diagnostic the spectrum of the demodulated in-loop PDH error signal measured by a (battery powered) floating FFT spectrum analyzer. The spectrum obtained after optimization of the lock performances is reported in Fig.~\ref{fig:spectrum} (black trace, c). The red trace (trace b) is the spectrum of the error signal when the laser is not resonant with cavity modes and represents our sensitivity limit. It is generated by the amplitude noise of the light and the main contribution to it is due to RAM of the PDH EOM. The blue trace (trace a) is the spectrum obtained with no light, that is the intrinsic noise of the detector and the electronic chain of the feedback loop. The traces are composed by three spectra for different Fourier frequency ranges normalized by their resolution bandwidth and the volt-to-hertz conversion factor is estimated from the PDH signal amplitude. Substantially at all frequencies above 1 Hz, the frequency noise is reduced under the Beta Line \cite{betaline}, that discriminates whether the noise spectral components either contribute (if above) or not (if below) to the laser linewidth. In the low frequency region, which is the most critical in ultranarrow laser systems, the frequency noise of the locked laser is reduced to the amplitude noise level. At higher frequencies the lock noise grows up to a maximum around $500$~kHz, which is our lock bandwidth, but still stays between $30$ and $40$~dB below the Beta Line level. In these in-loop spectra, the contribution of the cavity noise is not taken into account and the real laser linewidth can be determined only observing atomic transitions. \par

\section{ATOMIC SPECTROSCOPY}
The $578$~nm probe light is delivered to the atomic sample through a $10$~m long PM optical fiber contained in a protective tubing that provides insulation from air flows. No fiber noise cancellation scheme is implemented. Before the fiber, a second AOM (AOM2) operating at $40$~MHz, driven by a PC controlled synthesizer (Agilent 33522B), is used to compensate the linear aging drift of the ULE cavity, to finely tune the laser frequency and to rapidly switch light on and off (in cooperation with a fast mechanical shutter). The two AOMs have been chosen such that the total shift between the light sent to the atoms and the light delivered to ULE glass cavity covers exactly the frequency difference between the atomic and the cavity resonances. \par
We use $100$~ms long pulses of $\pi$ polarized light in order to perform spectroscopy on degenerate atomic samples of spin-polarized fermionic $^{173}$Yb. The atoms are confined in a deep Lamb-Dicke regime \cite{katori2003} in a 3D optical lattice at the magic wavelength $\lambda=759.35$~nm \cite{magic}.
\begin{figure}[t]
\centering
    \includegraphics[width=1\columnwidth]{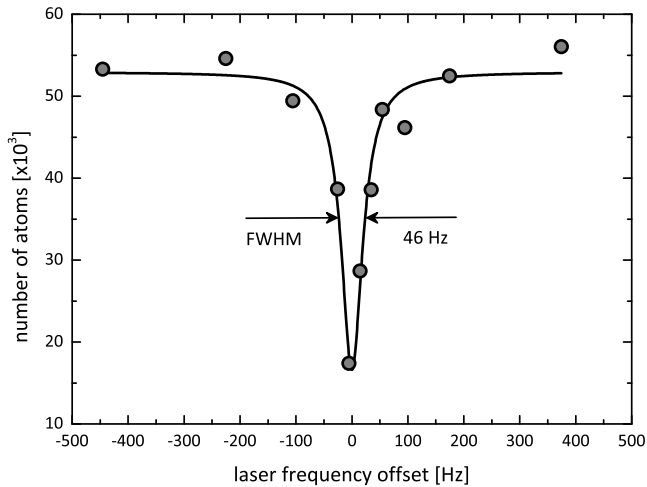}
\caption{Clock transition observed with a $0.3$~$\mu$W/mm$^2$ intensity and with $100$~ms probe time. Every point is acquired at the end of a $30$~s long experimental cycle for a final acquisition time of $5$~minutes. The FWHM linewidth is $46$~Hz.}\label{fig:riga}
\end{figure}  
A typical atomic transition spectrum is reported in Fig.~\ref{fig:riga}. It has a linewidth of $46$~Hz over a total acquisition time of $5$ minutes. The high frequency tails, which lay at $-50$~dBc/Hz, do not induce any visible feature. Narrower lines are not clearly distinguishable, as they are broadened due to a $30$~s long experimental cycle coupled to the residual drift of the laser, which is present even operating a feed-forward to cancel the cavity linear aging drift. This residual drift varies in a range of a few $0.1$~Hz/s during the day and we ascribe it to an imperfect thermal stabilization of the ULE cavity. At present, it represents the main limitation of the system. Improving the thermal stabilization of the cavity could be a first step to enhance the laser performances, but in order to perform experiments over timescales of some hours it is necessary to stabilize the laser to an absolute reference. For this purpose, we are planning to lock the laser to an absolute optical reference with a $10^{-14}$ stability at $1$~s generated at the Italian Metrological Institute (INRIM) and delivered at LENS in Florence through a $640$~km long fiber link \cite{fiberlink}.

\section{POWER BUDGET AND TRANSPORTABILITY}
Transportability is an important feature in the perspective of the realization of movable clocks to be employed outside of lab environment. To achieve it, not only a compact design of the laser system is required, but also a limited power consumption of its components. In our system, the laser setup (consisting of IR source, doubling cavity, fiber launch stages and AOMs) and the ULE cavity insulation enclosure fit on a $150$~cm~$\times$~$90$~cm optical table. All the electronics needed to run the system fit below the table frame, which is equipped with a set of wheels that allow the system to be easily moved. \par
On the power consumption side, in the present setup the laser source and the locking electronics are battery powered. Two $7$~Ah rechargeable acid lead batteries contained in the current driver supply the laser chip and two $24$~Ah batteries supply the locking electronics. The less sensitive electronic modules, including all temperature stabilization systems and the function generators that drive the EOM and AOMs are connected to the AC line supply. In the perspective of operating the entire system with a single battery supply, we estimate the total maximum power consumption of the system to be below $350$~W at the AC outlet, a value that can be easily sustained for several hours by a wide range of commercial UPS.

\section{CONCLUSIONS}
In this paper, we reported on the realization and characterization of a high power, ultranarrow laser system at $578$~nm for the excitation of Ytterbium atoms on the clock transition. We have described the infrared primary source and the doubling cavity used to generate the visible light, as well as the stabilization of the laser frequency to a high-finesse ULE cavity. We have also described in detail a low-cost, effective thermal insulation and stabilization system that allowed us to operate the ULE cavity at the mK level of stability around the zero CTE temperature.\par
Infrared ECDL, doubling cavity and all the optics, together with the ULE enclosure and the control electronics are placed on a single $150\times90$~cm$^2$ optical table, which can be easily moved and transported. The linewidth of the laser system is determined with atomic spectroscopy to be of the order of $40$~Hz on a timescale of $300$~s. The main limitation of the laser system is a residual drift of some $0.1$~Hz/s that can be reduced with a better thermal stabilization of the ULE cavity and locking the laser to an absolute optical reference. Moreover, the system is able to deliver up to $60$~mW of frequency-stabilized $578$~nm light, which represents a major advantage in applications with bosonic Ytterbium isotopes or for off-resonant manipulation on the clock transition.

\begin{acknowledgments}
We would like to thank D. Calonico and F. Levi for the supply of the ULE cavity and vacuum system and L. Lorini for early fruitful discussions. We are grateful to M. Siciliani de Cumis and P. Cancio Pastor for the help in characterizing the ULE cavity zero CTE temperature and K. Lauber for early suggestions on long base ECDLs. We also thank M. Inguscio and C. Sias for useful discussions and F. Schaefer for the contribution to the initial development of the system. This work was supported by EU FP7 Projects SIQS, MIUR Project PRIN2012 AQUASIM and ERC Advanced Grant DISQUA.
\end{acknowledgments}

\end{document}